\documentclass{mem}
\usepackage{natbib}\usepackage{txfonts}\usepackage{balance}
\usepackage{graphicx}
\usepackage[a4paper]{hyperref}
\idline{1}{1}
\begin{document}
\def\teff{$T\rm_{eff }$}
\def\kms{$\mathrm {km s}^{-1}$}

\title{
Formation and Evolution of Cataclysmic Variables
}

   \subtitle{}

\author{
H. \,Ritter
          }

  \offprints{H. Ritter}

\institute{
Max--Planck--Institut f\"ur Astrophysik, 
Karl--Schwarzschild--Str. 1,
D--85741 Garching, 
Germany, 
\email{hsr@mpa-garching.mpg.de}
}

\authorrunning{Ritter}

\titlerunning{Cataclysmic Variables}

\abstract{
This article summarizes the basic facts and ideas concerning
the formation and evolution of cataclysmic variables (CVs). It is
shown why the formation of CVs must involve huge
losses of mass and orbital angular momentum, very likely via a common
envelope evolution. A brief discussion of the principles of the
long-term evolution of semi-detached binaries follows. Finally a brief
sketch of CV evolution is given.

}
\maketitle{}

\section{Introduction}
Cataclysmic variables (CVs) are short-period semi-detached binary
systems in which a white dwarf (WD) primary accretes matter from a
low-mass companion star \citep{Warner1995}. CVs are intrinsically
variable and that on a wide range of time scales (from seconds to
$\gtrsim 10^6\,{\rm yr}$) and with a huge range of amplitudes (of up
to $10^6$ and possibly even more). The rich phenomenology of CV
variability which includes, among other things, phenomena like
flickering, dwarf nova and classical nova outbursts, can to a large
extent  be understood as either immediate or long-term consequences of
the mass transfer process. Interesting as all these phenomena are,
they are of no particular interest here. Rather, in the following I
shall concentrate on evolutionary aspects, i.e. on the formation and
evolution of CVs. Readers who are mainly interested in CVs as variable
stars should instead turn to the monographs by \citet{Warner1995} or
\citet{Hellier2002}.

\section{Very basic facts about CVs and stellar evolution}

\subsection{Generic properties of CVs} 
\label{CVproperties} 
From the perspective of stellar evolution a CV is a semi-detached
binary in which a WD primary of mass $M_1$ accretes from
a low-mass secondary star of mass $M_2$ which fills its critical Roche
lobe. From Roche geo\-metry it follows that the secondary's radius can
be written as $R_2 = a\, f_2(q)$. Here $a$ is the orbital separation, 
$q=M_1/M_2$ the mass ratio, and $f_2$ the fractional Roche radius of
the donor star. For typical values of $q$ found in
CVs, i.e. $1 \lesssim q \lesssim 10$,  Eqs.(\ref{f_2}) or (\ref{f_1}),
given below, yield $0.2 \lesssim f_2 \lesssim 0.4$.

In principle, the mass of the WD component can be anywhere between the
lowest possible value resulting from stellar evolution ($\sim 0.15
M_{\odot}$) and the Chandrasekhar mass $M_{\rm CH} \approx 1.4
M_{\odot}$. Observed masses are mostly in the range $0.5 M_{\odot}
\lesssim M_1 \lesssim 1 M_{\odot}$. As to the mass distribution
there are reasons to believe that intrinsically it is not unlike that  
of single WDs which have a mean  mass of $<M_{WD}> \approx 0.6
M_{\odot}$.  

From the observed mass transfer rates one can infer that mass transfer 
in CVs is stable. This, in turn, requires that the mass of the donor
is typically less than that of the WD component, i.e. $M_2 \lesssim
M_1$, or $q \gtrsim 1$, and thus that the donor is a low-mass star.
Observations show that in more than 95\% of all cases the donor star
is on the main sequence (MS), though not necessarily close to the zero
age main sequence (ZAMS). In rare cases the donor star is either a
giant, or a WD of very low-mass ($M_2 \lesssim 0.05 M_{\odot}$). 

For later comparison it is useful to keep in mind the resulting typical
system parameters of a CV with a MS donor: 
\begin{itemize} 
\item {\bf total mass:}~${\cal M} = M_1+M_2 \approx M_{\odot}$  
\vskip2mm
\item {\bf orbital separation:}~$a \approx {\rm few}~R_2 \approx
  R_{\odot}$  
\vskip2mm
\item {\bf orbital period:}~$80~{\rm min} \lesssim P_{\rm orb}
  \lesssim 10^{\rm h}$ 
\vskip2mm
\item {\bf orbital angular momentum:}
\begin{eqnarray}
\label{J_orb} 
J_{\rm orb} & = & G^{1/2} M_1~M_2~(M_1+ M_2)^{-1/2}~a^{1/2} \\ 
          & \approx & J_0 =  G^{1/2}~{M_\odot}^{3/2}~{R_\odot}^{1/2}\,. \nonumber
\end{eqnarray}  
\end{itemize}

\subsection{Evolution of single and binary stars}
In the following I summarize the basic facts which characterize single
star and binary evolution, and which are of relevance in the context
of our considerations. These facts are: 
\begin{enumerate} 
\item {\em Stars grow considerably as they age}\/~(by factors up to 
  $\gtrsim 10^2$). Because this growth is not strictly monotonic one
  can   distinguish distinct evolutionary phases during which a star
  grows. These phases are: 
\begin{itemize} 
\item central hydrogen burning, i.e. on the MS 
\item for intermediate mass and massive stars ($M\gtrsim 2.2 M_\odot$)
  the post-MS evolution towards He-ignition including the evolution
  through the Hertzsprung gap 
\item for low-mass stars ($M \lesssim 2.2 M_\odot$) evolution on the
  first giant branch up to the He-flash 
\item for low and intermediate mass stars ($M \lesssim 10 M_\odot$)
  evolution on the asymptotic giant branch (AGB)
\end{itemize}  
\item {\em The more massive a star, the faster it ages.} Stars on the
  main sequence obey a mass luminosity relation. On the upper MS
  ($1\,M_\odot \lesssim M\,\lesssim \,10\, M_\odot$) the luminosity $L$
  scales roughly as $L \propto M^{3.5}$. Hence the nuclear time scale
  is  $\tau_{\rm nuc} \propto M/L \propto M^{-2.5}$. The immediate
  consequence of this is that {\em of two stars with the same age (as
    in a binary) but different mass the more massive star grows
    faster, i.e. is the bigger of the two.} 
\item {\em In a binary the presence of a companion limits the size up
  to which a star can grow (Roche limit) without losing mass to its
  companion.} The maximum radii corresponding to the Roche limit are 
  the critical Roche radii $R_{1, \rm R} = a~f_1(q)$ and $R_{2, \rm R}
  = a~f_2(q)$ for respectively the primary and the secondary, where 
  $f_1(q) = f_2(1/q)$, and according to \citet{Paczynski1971} and
  \citet{Eggleton1983} for $1 \leqslant q \lesssim 10$ 
\begin{eqnarray} 
\label{f_2}
f_2(q) & \approx & 0.462\,(1+q)^{-1/3}, q \gtrsim 1.25 \\
\label{f_1}
f_1(q) & \approx & 0.38 + 0.2 \log q \approx q^{0.45}\,f_2(q)\,.
\end{eqnarray} 
As a consequence of this, stellar evolution in a binary of not too
large an orbital separation $a$ results sooner or later in the
formation of a so-called semi-detached binary in which the more
massive component reaches its Roche limit first and starts
transferring mass to its companion. 
\end{enumerate}

\subsection{Prerequisites for white dwarf formation}
WDs are the end product of the evolution of stars of low and
intermediate initial mass. Thereby the chemical composition of a WD 
reflects the evolutionary state of the star when it loses its
hydrogen-rich envelope. Depending on when this happens along the
evolution the result is either a WD consisting mainly of helium
(He-WD), of carbon and oxygen (CO-WD), or oxygen and neon (ONe-WD).  
\begin{itemize} 
\item {\bf He-WDs} result from the complete loss of the hydrogen-rich
  envelope of a low-mass star (with an initial mass $M_{\rm i}
  \lesssim 2.2 M_\odot$) on the first giant branch, i.e. before
  reaching the He-flash. Accordingly, the mass of He-WDs is in the
  range $0.15 M_\odot \lesssim M_{\rm He-WD} \leqslant M_{\rm He-Fl}$,
  where $M_{\rm He-Fl} \approx 0.45 - 0.50 M_\odot$ is the mass of
  the He core at the onset of the He-flash. Because wind mass loss of
  single stars on the first giant branch is not strong enough for
  complete envelope loss isolated He-WDs are not formed. However,
  they can result from mass transfer in a close binary (see e.g.\ 
  \citet{KKW1967}). 
\item {\bf CO-WDs} result from the complete loss of the hydrogen-rich
  envelope of intermediate mass stars on the AGB, i.e. before the
  onset of carbon burning. For single stars this happens if the
  initial mass is $M_{\rm i} \lesssim 6 - 8 M_\odot$. In binary stars
  this can happen for initial masses up to $\sim 10   M_\odot$.
  Accordingly, the resulting WD masses are in the range $M_{\rm He-Fl}
  \lesssim M_{\rm CO-WD} \leqslant M_{\rm C-ign}$, where $M_{\rm C-ign}     
  \approx 1.1 M_\odot$ is the core mass at the   onset of carbon
  ignition. 
\item {\bf ONe-WDs} originate from stars which undergo off-center
  carbon ignition and subsequent envelope loss during the so-called
  super-AGB phase. For single stars this is possible for initial
  masses in the range $9 M_\odot \lesssim M_{\rm i} \lesssim 10
  M_\odot$, whereas in binaries the mass range is $9 M_\odot \lesssim
    M_{\rm i} \lesssim 12 M_\odot$ (see e.g.\ \citet{Gil-Pons2001},
      \citet{Gil-Pons2003}). The resulting WDs have masses in the
      range $1.1 M_\odot \lesssim M_{\rm ONe-WD} \lesssim 1.38
      M_\odot$. 
\end{itemize} 
In the context of our considerations, one of the most important
properties of stars which have a degenerate core of mass $M_{\rm c}$
is that they obey by and large a core mass-luminosity relation 
${\cal L}(M_{\rm c})$, and to the extent that these stars have a
sufficiently massive hydrogen-rich envelope and thus are close to the 
Hayashi-line, also a core mass-radius relation ${\cal R}(M_{\rm c})$
(see e.g.\ \citet{Paczynski1970}, \citet{Kippenhahn1981},
\citet{Joss1987}). This relation shows that the radius of such a
star is a steeply increasing function of core mass and that, in
particular, AGB stars and stars on the super-AGB are very large with
radii of up to $\sim 10^3 R_\odot$. In other words: {\em the formation
of a WD requires a lot of space}, the more massive the WD the more
space. This is not a problem for single stars. But in a binary, as a
consequence of the Roche limit, the orbital separation $a$ sets an
upper limit to the mass of the WD that can be formed: $M_{\rm WD}
\lesssim {\cal R}^{-1}\left(a\,f_1(q)\right)$.

\subsection{Single star evolution versus binary star evolution} 
\label{*vs**-evol}
The task of calculating the structure and evolution of a single star
consists of solving a well-known set of differential equations with
appropriate boundary conditions and initial values
(e.g.\ \citet{Kippenhahn&Weigert1990}). 

For calculating the evolution of a binary system (or of one of its
components) the task is in principle the same as for single stars. The 
difference is that in a binary one has an additional boundary
condition which derives from the presence of the companion star,
i.e. from the Roche limit. 

Consider for simplicity a system consisting of a ``real `` star, say
the primary, and a point mass secondary. The simplest boundary
condition that one could impose in this case is that $R_1 \leqslant
a\,f_1(q)$. A more realistic approach would take into account that the
surface of a star is not arbitrarily sharp, but rather is
characterized by a finite scale height $H \ll R$ over which pressure,
density etc. drop off, by expressing the mass loss rate $-\dot M_1$ as
an explicit function of binary and stellar parameters
(e.g.\ \citet{Ritter1988}). What one finds is that $-\dot M_1$ is a
steeply increasing function of $(R_1-R_{1, \rm R})/H$ and that the
primary suffers significant mass loss as $R_1 \rightarrow R_{1, \rm R}$. 

The real problem when dealing with mass transfer consists of answering
two questions: 1.) Where does the mass lost from the donor go? and
2.) How much angular momentum does it take with it? On the formal
level this can be dealt with as follows: let us assume that a fraction
$\eta$ of the transferred mass is accreted by the secondary, i.e. 
\begin{equation} 
\label{defeta}
\dot M_2 = - \eta\, \dot M_1\,.
\end{equation} 
Accordingly, the mass loss rate from the system is $\dot{\cal M} =
(1-\eta)\,\dot M_1$. The angular momentum loss rate associated with
this mass loss can be written as 
\begin{equation}
\label{defnu} 
\dot
J_{\rm orb} = \nu\,\dot{\cal M} \, J_{\rm orb}/\cal M\,, 
\end{equation} 
where $\nu$ is a dimensionless factor measuring the angular momentum
leaving the system. 

What is known about the two parameters $\eta$ and $\nu$? In general
not more than 1.) $0 \leqslant \eta \leqslant 1$, and 2.) $\nu
\geqslant 0$. Otherwise $\eta$ and $\nu$ are (almost) free functions
of the problem. Therefore, {\em calculating binary evolution involves
  (at least) two almost free functions}. And the art of performing such
calculations very often consists of making creative use of this
freedom!
  
\subsection{Generic properties of CV progenitors}
\label{CVprogenitors}
We are now in a position to define the necessary criteria which a
binary consisting initially of two ZAMS stars of mass $M_{1, \rm i}$
and $M_{2, \rm i}$ has to meet in order to later become a CV which, at
the onset of mass transfer, i.e. at the beginning of its life as a CV,
consists of a WD of mass $M_{\rm WD}$ and a donor star of mass $M_2$. 
\begin{enumerate} 
\item {\em $M_{1, \rm i}$ has to have sufficient mass to allow for the
  formation of a WD of mass $M_{\rm WD}$.} 

  In theoretical calculations of the evolution of single stars with a
  fixed set of physical assumptions (such as initial chemical
  composition, equation of state, opacities, nuclear reaction rates,
  convection theory, wind mass loss, etc.) there is a one to one
  relation between the initial mass $M_{\rm i}$ and the mass 
  $M_{\rm f}$ of  the white dwarf produced. This relation is known as
  the {\em initial mass-final mass relation}, i.e. $M_{\rm WD} =
  M_{\rm f}(M_{\rm i})$. And, within the observational uncertainties,
  there is also ample observational evidence for this 
  $M_{\rm i}$-$M_{\rm f}$-relation (see e.g.\ \citet{Salaris2008} and
  references therein). 

  In binary evolution things are a little different: because mass
  transfer sets a premature end to the donor's nuclear evolution
  the mass of the resulting white dwarf is smaller than what single
  star evolution of the primary would yield, i.e. $M_{\rm WD} < 
  M_{\rm f}(M_{1, \rm i})$. In other words: for the formation of a WD
  of mass $M_{\rm WD}$ the necessary condition is $M_{1, \rm i} >
  {M_{\rm f}}^{-1}(M_{\rm WD})$. 
\item Because of the core mass-radius relation ${\cal R}(M_{\rm c})$
  which holds for the giant primary when it reaches its Roche limit,
  the initial separation of the binary must be 
  $a_{\rm i} = {\cal R}(M_{\rm WD})/f_1(q_{\rm i})$, where $q_{\rm i}$
  is the initial mass ratio. For this estimate of $a_{\rm i}$ we have
  implicitely assumed that after the onset of (the first) mass
  transfer   $M_{\rm WD} = {\rm const.}$
\item Finally for the secondary's mass we assume $M_{2, \rm i} =
  M_2$. A justification for this will be given below. 
\end{enumerate}  
Now, let us take typical parameters for a CV, say $M_{\rm WD} \approx 1
M_\odot$ and $M_2 \lesssim 1 M_\odot$, in order to see where this leads
us: with $M_{\rm WD} \approx 1 M_\odot$ it follows from the
$M_{\rm i}$-$M_{\rm f}$-relation that $M_{1, \rm i} \gtrsim 5 M_\odot$,
hence ${\cal M_{\rm i}} \gtrsim 6 M_\odot$, and from the core
mass-radius  relation ${\cal R}(M_{\rm WD}) \approx 10^3 R_\odot$, and
with $f_1(q_{\rm i}) \approx 0.5$, $a_{\rm i} \sim 2\,10^3 R_\odot$. 
Therefore,  the initial orbital angular momentum of the binary is 
\begin{eqnarray}
\label{J_orb,i}
J_{\rm orb, i}&=&J_0\,\left(\frac{M_{1, \rm i}}{M_\odot}\right) 
                   \left(\frac{M_{2, \rm i}}{M_\odot}\right) 
                  {\left(\frac{\cal M_{\rm i}}{M_\odot}\right)}^{-1/2} 
                  {\left(\frac{a_{\rm i}}{R_\odot}\right)}^{1/2}\\
           &\approx& 10^2\,J_0 \nonumber\, .
\end{eqnarray}
Comparing now the total mass and orbital angular momentum of a CV
(cf. Sect. \ref{CVproperties}) with the corresponding values of its
progenitor system we find that ${\cal M}_{\rm i}/{\cal M}_{\rm CV}
\approx 5 - 10$ and $J_{\rm orb,  i}/J_{\rm CV} \approx 10^2$. In other  
words: the formation of a CV invokes a binary evolution in which the
progenitor system has to lose $\sim 80\%-90\%$ of its initial mass
and up to $\sim 99\%$ of its initial orbital angular momentum
\citep{Ritter1976}, and that after the onset of mass transfer from the
primary.  

\section{Mass transfer and its consequences}
Since the primary of a CV progenitor does not stop growing when
approaching its Roche limit, onset of mass transfer is unavoidable.
And, because the subsequent formation of a CV involves huge losses of
mass and orbital angular momentum from the binary system, it is
necessary to examine the consequences of mass transfer for the ensuing
evolution in more detail. 

\subsection{Stability of mass transfer}
\label{stability} 
A detailed discussion of the stability of mass transfer is rather
complex and beyond the scope of this article. For this the reader is
referred to e.g.\ \citet{Ritter1988} or \citet{Ritter1996}. Here I
shall keep the presentation as simple as possible. 

Let us assume for the moment that the primary star has a sharp outer
boundary and that it has just reached its Roche limit, i.e. that $R_1
=  R_{1, \rm R}$. What happens if at that moment, which we denote by 
$t_0$, a small amount of mass $\delta m$ is taken away from the primary
and transferred to the secondary, i.e. if $M_1 \rightarrow M_1-\delta m$
and $M_2 \rightarrow M_2+\delta m$? As a consequence of this small
mass transfer, not only the mass ratio $q$ and the critical Roche radii 
$R_{1,\rm  R}$ and $R_{2, \rm R}$ will change but also the stellar
radii $R_1$  and $R_2$. Let us for the moment treat the secondary as a
point mass. Then we have to deal only with the radii $R_1(t>t_0)$ and
$R_{1, \rm R}(t>t_0)$. Thereby, three different situations can arise: 

\begin{enumerate}
\item $R_1(t>t_0) < R_{1, \rm R}(t>t_0)$: In this case {\em mass transfer
  is stable}, because after a small mass transfer $\delta m$ the donor
  underfills its critical Roche volume and mass transfer stops. 
\item $R_1(t>t_0) > R_{1, \rm R}(t>t_0)$: In this case {\em mass transfer
  is unstable}, because if $R_1(t>t_0) - R_{1, \rm R}(t>t_0) >0$~even
  more mass flows over. 
\item $R_1(t>t_0) = R_{1, \rm R}(t>t_0)$: In this case {\em mass transfer
  is marginally stable}. 
\end{enumerate} 
In order to decide which of the three above cases arises we must know
how $R_1$ and $R_{1, \rm R}$ react to mass transfer. For all practical
purposes $R_{1, \rm R}$ adjusts instantaneously (actually on the
orbital time scale) to changes in $M_1$, $M_2$ and $J_{\rm orb}$.
Although, in principle, calculating  $R_{1, \rm R}$ is straighforward,
for this it is still necessary to precisely specify where the
transferred mass goes and, if the system loses mass, how much angular
momentum it takes with it, i.e. one has to specify the parameters
$\eta$ and $\nu$. The change of $R_{1, \rm R}$ is conveniently
expressed in terms of the mass radius exponent 
\begin{equation}
\label{zeta_R} 
\zeta_{\rm R, 1} = {\left(\frac{\partial{\ln R_{1,\rm R}}}{\partial{ 
                 \ln M_1}}\right)}_*\,, 
\end{equation} 
where the subscript $*$ is a reminder that for its calculation $\eta$
and $\nu$ need to be specified. 

On the other hand, the reaction of the donor's radius $R_1$ to mass
loss is more complicated: besides hydrostatic equilibrium which
readjusts on the orbital time scale, mass loss disturbes also the
thermal equilibrium of a star. Therefore, its reaction depends on the
ratio of the mass loss time scale $\tau_{\rm M}$ to the time scale
$\tau_{\rm th}$ on which the star can readjust to thermal equilibrium.  
If $\tau_{\rm M}/\tau_{\rm th} \ll 1$ the star reacts essentially
adiabatically, and the radius change is expressed in terms of the
adiabatic mass radius exponent 
\begin{equation} 
\label{zeta_ad} 
\zeta_{\rm ad, 1} = {\left(\frac{\partial{\ln R_1}}{\partial{\ln M_1}}
               \right)}_{\rm ad}\,.
\end{equation} 
If, on the other hand, mass loss is very slow, i.e. $\tau_{\rm M}/
\tau_{\rm th} \gg 1$, the star has time to adjust to near thermal
equilibrium in which case the radius change is expressed by the
thermal equilibrium mass radius exponent 
\begin{equation} 
\label{zeta_th} 
\zeta_{\rm th, 1} = {\left(\frac{\partial{\ln R_1}}{\partial{\ln M_1}}
                  \right)}_{\rm th}\,.
\end{equation} 
Accordingly, there are two criteria for the stablity of mass transfer:  
\begin{enumerate} 
\item Mass transfer is {\em adiabatically} stable if 
\begin{equation}
\label{ad_stab}
\zeta_{\rm ad, 1} - \zeta_{\rm R, 1} > 0 
\end{equation}
\item Mass transfer is {\em thermally} stable if 
\begin{equation}
\label{th_stab}
\zeta_{\rm th, 1} - \zeta_{\rm R, 1} > 0\,. 
\end{equation}
\end{enumerate} 
What does all that mean for the CV progenitor system at the onset of
mass transfer? In order to tell one has to know the values of
$\zeta_{\rm R, 1}$, $\zeta_{\rm ad, 1}$, and $\zeta_{\rm th, 1}$.
Because $M_{1,\rm i}>M_{2, \rm i}$ one invariably finds that
$\zeta_{\rm R, 1} > 0$ even in the most favourable case where no
orbital angular momentum is lost. The values of $\zeta_{\rm ad, 1}$,
and $\zeta_{\rm th, 1}$, on the other hand, depend on the internal
structure of the star in question. In our case the donor is a star
with a degenerate core and a deep outer convective envelope. For such
stars one typically finds $-1/3 \lesssim \zeta_{\rm  ad}  \lesssim 0$
and $\zeta_{\rm th} \lesssim 0$ \citep{HW87}. Taken together this
means that mass transfer in such a system is adiabatically and
thermally unstable. And as a consequence of the adiabatic instability
mass transfer quickly accelerates to the point where the mass transfer
rate reaches values of order of $-{\dot M}_{1, \rm ad}
\sim M_1/\tau_{\rm conv} \sim M_\odot{\rm yr}^{-1}$, where 
$\tau_{\rm conv}  \sim {\rm yr}$ is the convective turnover time scale
\citep{PS72}.

\subsection{Fast accretion onto a main sequence star} 
So far we have treated the MS secondary as a point mass. Whereas
before the onset of mass transfer this is an adequate approximation,
this is not always true afterwards. Numerical calculations (e.g.\
\citet{K&M-H77}, \citet{NMNS77}) show that the low-mass secondary,
exposed to the prodigious mass inflow rates associated with the
adiabatic mass transfer instability, starts expanding rapidly to giant 
dimensions. The reason for this behaviour is that the thermal time
scale of the accreted envelope around the secondary is much longer
than the mass accumulation time. As a consequence, the accreted matter
can not cool efficiently and, therefore, forms a deep and very
extended convective envelope of high entropy material around the
secondary. The star thus attains a structure similar to that of a
giant/AGB star which, however, derives its luminosity mainly from
accretion rather than from nuclear burning. 

\subsection{Formation of a common envelope} 
The situation of a CV progenitor at the onset of mass transfer can now
be characterized as follows: because mass transfer occurs from the
more massive star, the orbital separation $a$ as well as the critical
Roche radii $R_{1, \rm R}$ and $R_{2, \rm R}$ shrink. At the same
time, the  mass losing donor star has the tendency to expand (negative
$\zeta_{\rm ad}$ and $\zeta_{\rm th}$). But forced by dynamical
constraints to essentially follow $R_{1, \rm R}$ the donor must lose
mass at rates approaching $\sim M_\odot{\rm yr}^{-1}$. And the
secondary, in turn, exposed to such enormous accretion rates, reacts by
rapid expansion. The consequence of all this is that within a very
short time after the onset of mass transfer the system evolves into
deep contact. An attempt to model this very complicated process has
been made by \citet{Webbink1979}. Accordingly, the immediate result of
this evolution can then be roughly chracterized as follows: A binary
system consisting of the primary's core (the future WD) of mass
$M_{\rm c}$ and the original secondary of mass $M_{2, \rm i}$ finds
itself deeply immersed in a common envelope (CE) of mass $M_{\rm CE} = 
M_{1, \rm i}-M_{\rm c}$ and a size which must be of order of or even 
larger than the radius given by the core mass-radius relation, i.e.
$R_{\rm CE} \gtrsim {\cal R} (M_{\rm c})$. 

\section{Common envelope evolution and CV formation}
Common envelope evolution is the name of a generic process which
arises as a consequence of dynamical time scale mass transfer and as
a result of which a detached short-period binary is formed in which
one of its components is the core of the former primary (in our
case a pre-WD). Because of its importance for the formation of all
sorts of compact binaries the subject has generated a vast
literature. For lack of space I am unable to give a detailed review
here. Rather I shall concentrate on sketching a few key aspects of this
process and for more details refer the reader to recent reviews by 
\citet{Taam&Sandquist2000} and \citet{Webbink2008}.   

\subsection{The Darwin instability} 
Let us now consider the following idealized situation: a binary
consisting of the original primary's core of mass $M_{\rm c}$ and the
secondary of mass $M_2$ with orbital separation $a$ and orbital
frequency $\omega_{**}$ is embeded in an envelope of mass $M_{\rm E}$,
radius $R_{\rm E}$, moment of inertia $I_{\rm E}$ which is in solid body
rotation with an angular frequency $\Omega_{\rm E}$. If $\omega_{**} > 
\Omega_{\rm E}$ tidal interaction and friction between the binary and
envelope lead to energy dissipation and angular momentum transport
from the binary to the envelope with ${\dot J}_{**}=-{\dot J}_{\rm  E}
< 0$. As a consequence, the envelope, initially rotating slower than
the binary, is spun up. But according to Kepler's third law also the
binary's orbital frequency increases due to the loss of orbital
angular momentum. The question of interest is thus whether through
this spin-up the difference $\omega_{**}-\Omega_{\rm E}$ increases or
decreases. 

If $\omega_{**}-\Omega_{\rm E} > 0$ and
$\dot\omega_{**}-\dot\Omega_{\rm E} <0$  
the envelope is synchronized, i.e. $\Omega_{\rm E} \rightarrow
\omega_{**}$. 

If, on the other hand, $\omega_{**}-\Omega_{\rm E} > 0$ and
$\dot\omega_{**} -\dot\Omega_{\rm E} >0$, runaway friction results,
and the binary spirals in. The condition for this to happen is easily
derived: the binary's orbital angular momentum is 
\begin{eqnarray} 
\label{J_**}
J_{**}&=&G^{2/3} \frac{M_{\rm c} M_2}
        {(M_{\rm c}+M_2)^{1/3}}\,{\omega_{**}}^{-1/3} \\
     &=&I_{**} \omega_{**}\,, 
\end{eqnarray} 
where 
\begin{equation} 
\label{I_**}
I_{**} = \frac{M_{\rm c} M_2}{M_{\rm c}+M_2}\,a^2
\end{equation} 
is the orbital moment of inertia. The envelope's spin angular momentum 
is 
\begin{equation} 
\label{J_E} 
J_{\rm E} = I_{\rm E}\,\,\Omega_{\rm E}\,.
\end{equation}  
With (\ref{J_**}) and (\ref{J_E}) angular momentum conservation, i.e. 
$\dot J_{**} + \dot J_{\rm E} =0$, yields 
\begin{equation} 
\label{dotomegadiff} 
\dot \omega_{**} - \dot \Omega_{\rm E} = \dot \omega_{**} 
  \left({1 - \frac{1}{3} \frac{I_{**}}{I_{\rm E}}}\right)\,.
\end{equation} 
From (\ref{dotomegadiff}) it is seen that the envelope can be
synchronized only if $I_{\rm E} < {1/3} I_{**}$. If, on the other hand, 
\begin{equation} 
\label{Darwininst}  
I_{\rm E} > \frac{1}{3}\,I_{**}
\end{equation} 
the envelope cannot be synchronized and spiral-in of the binary is
unavoidable. The impossibility of synchronizing the envelope results
from a variant of an instability which is actually long known:
discovered by \citet{Darwin1879}, though in a different context, it is 
commonly called Darwin instability.  

Whether the Darwin instability is of relevance for our problem, i.e. 
whether the criterion (\ref{Darwininst}) is met with the formation of a
CE after the onset of adiabatically unstable mass transfer, needs of
course first to be checked. Since adequate model calculations of the
formation of a CE are still not feasible, simple estimates must
do. And these indicate indeed that for typical parameters of CV 
progenitor systems the forming CE systems are Darwin unstable.

\subsection{Common envelope evolution} 
Despite decades of heroic efforts to model common envelope evolution,
for a review see e.g.\ \citet{Taam&Sandquist2000}, to this day it has
not yet been possible to follow such an evolution from its beginning
to its end with really adequate numerical computations. Therefore, it 
is still not possible for a given set of initial parameters to 
reliably predict the outcome of common envelope evolution. The
expectation is that in many, but not necessarily all cases the
frictional energy release will unbind the CE and leave a close binary
consisting of the former primary's degenerate core and the secondary.

Clearly the ejection of the CE requires the release of the envelope's
binding energy in a sufficiently short time, i.e. that the time scale
of the spiral-in is short. However, there are limits to how short the
spiral-in can be. From simplified one-dimensional hydrostatic model
calculations \citet{M&M-H1979} found that there is a negative feedback
between the frictional energy release and the resulting radiation
pressure. An estimate of the duration of the spiral-in is obtained
from the argument that because of this feedback the frictional
luminosity $L_{\rm frict}$ can not exceed the Eddington luminosity  
\begin{equation}
\label{L_Edd} 
L_{\rm Edd} = \frac{4\,\pi\,G\,c\,\cal M}{\kappa_{\rm es}}
\end{equation} 
by much. Here $\kappa_{\rm es}$ is the electron scattering opacity.
The evolution of the binary with masses $M_{\rm c}$ and $M_2$ from an 
initial separation $a_{\rm i}$ to a final separation $a_{\rm f} \ll
a_{\rm i}$ releases the orbital binding energy 
\begin{equation}
\label{Delta_E_B**}
\Delta E_{\rm B} \approx \frac{G\,M_{\rm c}\,M_2}{2a_{\rm f}}\,.
\end{equation}
This yields a rough estimate of the spiraling-in time scale 
\begin{eqnarray}
\label{tau_CE} 
\tau_{\rm CE} &\approx& \frac{\Delta E_{\rm B}}{L_{\rm frict}}
             \gtrsim  \frac{\Delta E_{\rm B}}{L_{\rm Edd}} \\
            &\gtrsim& 400{\rm yr}\,
                      \frac{M_{\rm c}\,M_2}{(M_{\rm c}+M_2)\,M_\odot}\,
                      \frac{a_{\rm f}}{R_\odot}\,.
\end{eqnarray} 
Thus for the typical parameters of a CV (see Sect.\ref{CVproperties})
$\tau_{\rm CE}$ is very short, so short indeed that the secondary star
has no time to accrete a significant amount of mass during the CE
phase \citep{H&T1991}. This is the a posteriori justification for our
assumption in Sect. \ref{CVprogenitors} that $M_{2,\rm i} = M_{2,\rm f}$.                         
            
Because of the short duration of CE evolution chances of observing
a binary system during this phase are extremely small, apart from the
fact that it is not even quite clear what to look for. Worse, the 
spiraling-in binary is hidden from view as long as it is inside the
CE. In view of our limited theoretical understanding of CE evolution in
general and the ejection of the CE in particular, and the fact that
this process is virtually unobservable, one has to ask why we can be
sure that CE evolution really happens as described above. Beyond all
the uncertainties, the concept of CE evolution does make at least one 
prediction that is testable: at the end of the CE process, if the
envelope is ejected, we expect a binary inside the now more or less 
transparent envelope. And in this binary the primary's degenerate core 
emerges as a very hot pre-WD which, in turn, ionizes the surrounding
gas, thereby transforming the ejected CE into a planetary nebula. The
concept of CE evolution thus implies the existence of planetary
nebulae with short-period binary central stars. And indeed, such
objects are obeserved: currently we know of $\sim 20$ short-period
binary central stars of planetary nebulae (see
e.g.\ \citet{DeMarco2008} and \citet{RKcat}). 

\subsection{Formal treatment of the CE phase} 
CE evolution, if it ends with the ejection of the CE, transforms a
binary with initial parameters ($M_{1,\rm i}, M_{2,\rm i}, a_{\rm i}$)
to one with final parameters ($M_{1,\rm f}, M_{2,\rm f}, a_{\rm f}$).
With current theory it is not possible to precisely link these two
sets of parameters. Therefore, in evolutionary studies and population
synthesis calculations of compact binaries (e.g.\ \citet{deKool1990},
\citet{deKool1992}, \citet{dK&R1993}, \citet{Politano1996},
\citet{Politano2004}, \citet{Politano2007}), CE evolution is usually
dealt with by means of a simple estimate \citep{Webbink1984} which
derives from the assumption that a fraction $\alpha_{\rm CE} \lesssim
1$ of the binary's binding energy which is released in the
spiraling-in process, $\Delta E_{\rm B, **}$, is used to unbind the
CE.  

Using $M_{1,\rm f} = M_{\rm c, i} = M_{\rm c}$, $M_{2,\rm f} =
M_{2,\rm i} = M_2$ we have 
\begin{equation}
\label{Delta_EB}
\Delta E_{\rm B, **} = \frac{G\,M_{\rm c}\,M_2}{2}\,
                    \left(\frac{1}{a_{\rm i}}
                    -\frac{1}{a_{\rm f}}\right) \,.
\end{equation} 
On the other hand, the binding energy of the CE can be written as 
\begin{equation} 
\label{EB_CE} 
E_{\rm B, CE} = - \frac{G\,M_{1,\rm i}\,M_{\rm CE}}
                     {\lambda \,R_{1,\rm i}}\,,
\end{equation} 
where $M_{\rm CE} = M_{1,\rm i} - M_{\rm c}$ is the mass and 
$R_{1,\rm i} = a_{\rm i}\,f_1(q_{\rm i})$ the radius of the CE, and
$\lambda$ a dimensionless factor which can be determined from
stellar structure calculations provided one knows exactly where 
the mass cut between core and envelope is. Unfortunately it turns out
that $\lambda$ depends rather sensitively on this \citep{T&D2001}.
The CE criterion, namely that 
\begin{equation} 
\label{CEcriterion} 
E_{\rm B, CE} = \alpha \, \Delta E_{\rm B, **}   
\end{equation} 
is then equvalent to 
\begin{equation} 
\label{a_f}
a_{\rm f} = a_{\rm i}\,{\left\{\frac{2\,M_{1,\rm i}\,M_{\rm CE}}
          {\alpha_{\rm CE}\,\lambda \,M_{\rm c}\,M_2\,f_1(q_{\rm i})} 
          - \frac{M_{1,\rm i}}{M_{\rm c}}\right\}}^{-1}\,.
\end{equation}   
Eq. (\ref{a_f}) provides the formal link between the pre-CE
and the post-CE binary parameters. As can be seen from Eq. (\ref{a_f})
when dealing with CE evolution in this way one introduces essentially
one free parameter, namely $\alpha_{\rm CE}\,\lambda$ (per CE phase).
Since we do not have any a priori knowledge about $\alpha_{\rm CE}$
and since also $\lambda$ is not really well known, the degree of
uncertainty introduced via $\alpha_{\rm CE}\, \lambda$ is quite
considerable. 

Several recent investgations of binary evolution involving CE
evolution have come to the conclusion that the energy criterion
(\ref{CEcriterion}) is not always adequate and that in addition to the 
orbital binding energy possibly also other sources of energy such as 
the ionization energy have to be taken into account. For a
comprehensive discussion of this point see \citet{Webbink2008}. 

\subsection{Evolution of post-common envelope binaries} 
\label{preCVevol}
The ejection of the CE leaves a detached short-period binary inside a
planetary nebula which is excited by the hot pre-WD component. Once
the planetary nebula disappears, either because it dissolves or
because of lack of ionizing radiation from the pre-WD, what remains is
a binary consisting of a WD and an essentially unevolved
companion. And because the lifetime of a typical planetary nebula of
$\sim 10^4\,{\rm yr}$ is much shorter than the lifetime of a typical
post-CE binary in the detached phase, the intrinsic number of detached
post-CE systems lacking a visible planetary nebula must be vastly
larger than that of post-CE systems with a planetary nebula. And
although such systems are intrinsically rather faint (both the WD and
its low-mass companion are faint), because of their rather high space
density quite a number of such systems are known (currently $\gtrsim
50$, see \citet{RKcat} for a compilation). They are collectively
refrerred to as {\em precataclysmic binaries}, hereafter pre-CVs. 

In the following, we need to discuss two questions: 1) how does a
detached pre-CV become semi-detached, i.e. a CV, and 2) whether with
the onset of mass transfer all pre-CVs really become CVs or perhaps
follow a totally different evolutionary path. 

Since in a detached system the future donor star underfills its
Roche lobe, mass transfer can only be initiated if either the donor
star grows (as a consequence of nuclear evolution) or if the orbital
separation shrinks as a consequence of orbital angular momentum loss
(AML). Which of the two possibilities is relevant for a particular
binary system depends on the ratio of the nuclear time scale 
\begin{equation}
\label{tau_nuc} 
\tau_{\rm nuc, 2} = {\left(\frac{\partial t}{\partial \ln
    R_2}\right)}_{\rm nuc}
\end{equation} 
on which the star grows to the AML time scale 
\begin{equation} 
\label{tau_J} 
\tau_{\rm J} = -\,{\left(\frac{\partial t}{\partial \ln J_{\rm orb}}\right)} 
           = -2\, \left(\frac{\partial t}{\partial \ln a}\right) 
\end{equation} 
on which the orbital separation $a$ shrinks. 

If $\tau_{\rm J} < 2\, \tau_{\rm nuc, 2}$ mass transfer is initiated by
AML, otherwise by nuclear evolution. The typical future donor star of
a pre-CV is a low-mass MS star. Thus $\tau_{\rm nuc, 2} >
10^9\, {\rm yr}$. AML in such binaries results either from the
emission of gravitational waves \citep{KM&G1962} or from {\em magnetic 
  braking}, i.e. a magnetically coupled stellar wind from the tidally
locked companion. In typical pre-CV systems AML is dominated by
magnetic braking. Unfortunately, for that case there is as yet no
theory which would allow computation of $\dot J_{\rm orb}$ from first
principles. Again, simple semi-empirical estimates
(e.g.\ \citet{V&Z1981}) or simplified theoretical approaches (e.g.\
\citet{M&S1987}) must do. For the typical pre-CV with a low-mass MS
companion, these estimates yield $\tau_{\rm J} \sim 10^8\,{\rm yr}$.   
Thus, for such systems mass transfer is typically initiated via AML
(see e.g.\ \citet{Ritter1986}, \citet{S&G2003}). But the simple fact
that we do observe a number of long-period CVs with a giant donor
shows that mass transfer can also be initiated by nuclear evolution of
the future donor star. However, the fraction of pre-CV systems ending
up with a giant donor is small and, unfortunately, strongly
model-dependent 
\citep{deKool1992}. 

When the secondary reaches its Roche limit and mass transfer sets in
stability of mass transfer becomes again an issue. Whether mass
transfer is stable depends on whether the criteria which we had
derived in Sect. \ref{stability}, but now applied to the secondary
star, are fulfilled. Why is this important? Observations and
theoretical arguments show that in the vast majority of CVs mass
transfer is thermally and adiabatically stable. In other words: only
those pre-CVs for which $\zeta_{\rm ad, 2} - \zeta_{\rm R, 2} > 0$ and
$\zeta_{\rm th, 2} - \zeta_{\rm R, 2} > 0$ can directly become CVs. What   
happens to the rest? That depends mainly on the evolutionary status
of the donor and the binary's mass ratio. If we distinguish for
simplicity MS stars and giants as possible donor stars, then the
following cases can arise: 
\begin{enumerate} 
\item MS donor, mass transfer thermally and adiabatically stable
  $\rightarrow$ short-period CV ($P_{\rm orb} \lesssim 0.5\,{\rm d}$)
  with an unevolved donor. 
\item MS donor, mass transfer adiabatically stable but thermally
  unstable $\rightarrow$ thermal time scale mass transfer, WD with
  stationary hydrogen burning, system appears as a supersoft X-ray
  source (see e.g.\ \citet{vdHBNR1992}, \citet{SKKWZ2002})
  $\rightarrow$ CV with an artificially evolved MS donor. 
\item MS donor, mass transfer adiabatically unstable $\rightarrow$
  very high mass transfer rates, second common envelope?,
  coalescence?  
\item giant donor, mass transfer thermally and adiabatically stable
  $\rightarrow$ long-period CV ($P_{\rm orb} \gtrsim 1\,{\rm d}$). 
\item giant donor, mass transfer either thermally or adiabatically
  unstable $\rightarrow$ very high mass transfer rates, second common 
  envelope?, formation of an ultrashort-period detached WD+WD binary? 
\end{enumerate}

\section{CV evolution} 
CV evolution is a complex subject. Yet, because of space constraints,
here I can only present a brief outline of this topic. For readers
wishing to learn more about it the reviews by \citet{King1988} and
\citet{Ritter1996} are a good starting point.  

\subsection{Mass transfer in semi-detached binaries}
If mass transfer in a binary is thermally and adiabatically stable, as
in the majority of CVs, no mass transfer occurs unless some external
force drives it. And in CVs the driving agents are the same as in
pre-CVs (cf. Sect. \ref{preCVevol}), i.e. AML and nuclear evolution of
the donor. Furthermore, if mass transfer is stable and the strength of
the driving changes only on long time scales, mass transfer will be
essentially stationary. In that case the donor's radius $R_2$ and its
Roche radius $R_{2,\rm R}$ are equal to within very few atmospheric
scale heights $H \ll R_2$ \citep{Ritter1988}. Thus, to a very good
accuracy we must have $\dot R_2 = \dot R_{2,\rm R}$, or, using $R_2 =
R_{2,\rm R}$,  
\begin{equation} 
\label{stationarity} 
\frac{d \ln R_2}{dt} = \frac{d \ln R_{2,\rm R}}{dt}\,.
\end{equation} 
Now, the donor's radius can change because of mass loss, nuclear
evolution, and thermal readjustment. As mentioned earlier
(Sect. \ref{stability}) mass loss (if nothing else) drives a star out
of thermal equilibrium. If mass loss were stopped the star evolved
back towards thermal equilibrium, thereby changing it radius initially
at a relative rate 
\begin{equation} 
\label{thermalrelax}
\left(\frac{\partial \ln R_2}{\partial t}\right)_{\rm th} =
                                                \frac{1}{\tau_{\rm th,2}}\,,
\end{equation} 
where $\tau_{\rm th,2}$ is the thermal time scale. Thus the rate of
change of $R_2$ can be decomposed as follows:  
\begin{equation} 
\label{dlnR2/dt} 
\frac{d \ln R_2}{dt} = \frac{\dot M_2}{M_2}\,\zeta_{\rm ad, 2}
                         +\frac{1}{\tau_{\rm th,2}} 
                         +\frac{1}{\tau_{\rm nuc,2}}
\end{equation} 
On the other hand, the donor's Roche radius can change because of mass 
transfer and AML. With (\ref{tau_J}) we have 
\begin{equation} 
\label{dlnR2R/dt} 
\frac{d \ln R_{2,\rm R}}{dt} = \frac{\dot M_2}{M_2}\,\zeta_{\rm R, 2}
                              -\frac{2}{\tau_{\rm J}}\,.
\end{equation}
Eqs. (\ref{stationarity}), (\ref{dlnR2/dt}), and (\ref{dlnR2R/dt})
finally yield the mass transfer rate 
\begin{equation} 
\label{dotM2}
-\dot M_2 = \frac{1}{\zeta_{\rm ad, 2}-\zeta_{\rm R, 2}}\,
           \left(\frac{1}{\tau_{\rm th,2}}
              +\frac{1}{\tau_{\rm nuc,2}} 
              +\frac{2}{\tau_{\rm J}}\right)\,.
\end{equation} 
If mass transfer is sufficienly slow such that the donor remains close
to thermal equilibrium, its radius changes according to  
\begin{equation} 
\label{dlnR2/dt_th} 
\frac{d \ln R_2}{dt} = \frac{\dot M_2}{M_2}\,\zeta_{\rm th, 2}
                         \,+\,\frac{1}{\tau_{\rm nuc,2}}\,,
\end{equation} 
and together with (\ref{stationarity}) and (\ref{dlnR2R/dt}) we can
write 
\begin{equation} 
\label{dotM2_th} 
-\dot M_2 = \frac{1}{\zeta_{\rm th, 2}-\zeta_{\rm R, 2}}\,
            \left(\frac{1}{\tau_{\rm nuc,2}}\, 
           +\,\frac{2}{\tau_{\rm J}}\right)\,.
\end{equation} 
From what has been said so far, it is easily seen that for the sign of
the mass transfer rate to be correct, i.e. for $-\dot M_2>0$, the
denominator in (\ref{dotM2}) and (\ref{dotM2_th}) must be positive,
i.e. that 
\begin{equation}
\label{adstab}
\zeta_{\rm ad, 2}-\zeta_{\rm R, 2}>0
\end{equation} 
 and  
\begin{equation}
\label{thstab}
\zeta_{\rm th, 2}-\zeta_{\rm R, 2}>0\,. 
\end{equation} 
With Eqs. (\ref{adstab}) and (\ref{thstab}) we have thus recovered the
stability criteria for mass transfer. What this implies is that for
mass transfer to be stationary it has also to be stable.

\subsection{Computing the evolution of a semi-detached binary} 
Can we use Eqs. (\ref{dotM2}) or (\ref{dotM2_th}) for calculating the 
evolution of a semi-detached binary? Unfortunately, this is in general
not the case. The virtue of Eqs. (\ref{dotM2}) or (\ref{dotM2_th}) and
the reason why we have derived them here is that they show clearly how
the long-term evolution of a semi-detached binary works: mass transfer
must be stable and be driven by some mechanism. The obvious ones are
the growth of the donor star due to nuclear evolution or AML which
shrinks the binary. A less obvious driving agent is the growth of the
donor star as a consequence of thermal relaxation
(cf. (\ref{dotM2})). However, thermal relaxation, itself mainly being 
a consequence of mass loss, cannot maintain mass transfer for times 
long compared to $\tau_{\rm th}$ without external driving by one of the 
other mechanisms. 

The reason why we cannot use Eqs. (\ref{dotM2}) or (\ref{dotM2_th})
for evolutionary computations is that most of the quantities appearing
in these equations are not explicitly known. In partcular, 
$\zeta_{\rm ad}$, $\zeta_{\rm th}$, $\tau_{\rm nuc}$, and $\tau_{\rm th}$ 
require knowledge of the complete internal structure of the donor
star, i.e. nothing less than the whole past history of the binary
system. Worse, even if all that were known, the above quantities can
only be determined numerically. Furthermore, computing $\zeta_{\rm R}$
requires specification of $\nu$ and $\eta$ (Eqs. (\ref{defnu}) and
(\ref{defeta})). Finally, apart from gravitational radiation, the AML
rate is not well known and in some cases only given as an implicit
function of binary parameters \citep{M&S1987}. Even more exotic
effects such as irradiation of the donor star or the accretion disc
can strongly affect the quantities appearing in (\ref{dotM2}) or
(\ref{dotM2_th}) (see e.g.\ \citet{Ritter1996}, \citet{B&R2004}, or
\citet{Ritter2008} for more). 

Application of Eqs. (\ref{dotM2}) or (\ref{dotM2_th}) for evolutionary
computations is therefore limited to cases where the donor star can
either be approximated by a particularly simple stellar model, e.g. by
a polytrope \citep{RJ&W1982}, a bipolytrope (e.g.\ \citet{RVJ1983},
\citet{Kolb&Ritter1992}), or where stellar structure data determined
beforehand from single star evolution can be used (\citet{WR&S1983},
\citet{Ritter1999}). 

In general, such simplifications are un\-satis\-factory. For a more
realistic simulation the full stellar structure problem must be solved
as described in Sect. \ref{*vs**-evol}. Stellar evolution is an
initial value problem. So for setting up a simulation of a CV
evolution one has first to decide at which moment of the evolution to
start the calculation, e.g. at the onset of mass transfer from the
secondary, and then to specify at least the masses of the components
and the internal structure, i.e. the evolutionary status of the donor
star, but as the case may be also the structure of the accreting
WD. Furthermore, one has to adopt values or prescriptions for $\nu$ and
$\eta$, and finally to decide what to do about AML, in particular
about magnetic braking, i.e. which of the various prescriptions
available in the lierature (e.g.\ \citet{V&Z1981}, \citet{M&S1987}) to
use. When everything is set up calculating the evolution is in the
simplest case just a single star evolution for the donor star with
variable mass where the mass loss rate is an eigenvalue of the problem
and is determined by the additional outer boundary condition , e.g. by
$R_2 \leqslant R_{2, \rm R}$.

\subsection{A sketch of CV evolution} 
The orbital period $P_{\rm orb}$ is the only physical quantity which
is known with some precision for a large number of CVs, currently for
over 700 objects \citep{RKcat}. Reliable masses, on the other hand,
are known, if at all, only for a very small minority of CVs. Therefore,
much of the work on CV evolution in the past 30 years has concentrated
on understanding the observed period distribution of CVs. Broadly
speaking, this distribution is bimodal with $\sim 45\%$ of the objects
having periods in the range $3^{\rm h} \lesssim P_{\rm orb} \lesssim
16^{\rm h}$, another $\sim 45 \%$ with $80\,{\rm min} \lesssim
P_{\rm orb}  \lesssim 3^{\rm h}$, and the remaining $\sim 10 \%$ with 
$2^{\rm h} \lesssim P_{\rm orb}  \lesssim 3^{\rm h}$. The dearth of objects 
in the period interval $2^{\rm h} \lesssim P_{\rm orb} \lesssim 3^{\rm h}$  
is known in the literature as the {\em period   gap}.  

The maximum period of $\sim 16^{\rm h}$ is easily understood as a
consequence of the facts that 1) the donor is a MS star, 2) the
mass of the WD is $M_{\rm WD} < M_{\rm CH} \approx 1.4 M_\odot$ and 3) 
mass transfer must be stable. 

The minimum period of $\sim 80\,{\rm min}$, in turn, is at least
qualitatively understood as a consequence of mass transfer from
a hydrogen-rich donor which is mainly driven by gravitational
radiation (\citet{P&S1981}, \citet{P&S1983}, \citet{RJ&W1982}).
Because of mass loss, of the order of a few $10^{-11} M_\odot 
{\rm yr}^{-1}$, the donor star  becomes more and more degenerate when
$M_2 \lesssim 0.1 M_\odot$ and its structure changes from that of a
low-mass MS star to that of a brown dwarf. Thereby its effective mass
radius exponent $\zeta_{\rm eff, 2} = d \ln R_2/d \ln M_2$  changes
from $\sim 0.8$ on the MS to $-1/3$. $P_{\rm orb}$ is minimal when
$\zeta_{\rm eff,2} = +1/3$. Whether mass transfer near the period
minimum is really driven by gravitational radiation only is currently
under dispute because of the mismatch between the corresponding
theoretical prediction for the minimum period of $\sim 70\,{\rm min}$
and the observed value of $\sim 77\,{\rm min}$ (see
e.g.\ \citet{RBK&R2002} or \citet{B&K2003} for a discussion). 

The period gap is more difficult to account for. Over the years a
number of different hypotheses have been put forward to explain it. For
lack of space I cannot review them all here. Rather I shall
concentrate on the one hypothesis (\citet{Spruit&Ritter1983},
\citet{RVJ1983}) which, in my view, still provides the most plausible
explanation for what we see, and which is known in the literature  as
the {\em disrupted (magnetic) braking} hypothesis. It postulates that
as long as the donor star has a radiative core ``magnetic braking'' is
effective and CV evolution is driven by a high AML rate due to
``magnetic braking'' and gravitational radiation, but that, as soon as
the donor star becomes fully convective, ``magnetic braking'' becomes
ineffective and thus the evolution is driven by AML from gravitational
radiation only. In the following I shall try to explain step by step
how the gap arises in the framework of this hypothesis. 

First it is important to note that the evolution of CVs with a MS
donor driven by AML leads from longer to shorter orbital periods. And
MS donor stars with a mass $\lesssim 1 M_\odot$ have a convective
envelope and a radiative core. With decreasing mass, i.e. 
$P_{\rm orb}$, the mass of the radiative core shrinks until at a
particular mass $M_{2,\rm conv}$, i.e. orbital period $P_{\rm orb} =
P_{\rm u}$, the donor becomes fully convective. According to the above
hypothesis, at this point the AML rate drops from a high value which
is mainly due to ``magnetic braking'' to a small value due to 
graviational radiation only. 

If ``magnetic braking'' is sufficiently strong, then for periods
$> P_{\rm u}$ mass loss from the donor occurs on a timescale much
shorter than its thermal time scale. As a result the donor is
significantly driven out of thermal equilibrium and, therefore,
oversized compared to its thermal equilibrium radius, i.e. 
$R_2(P_{\rm orb}> P_{\rm u})> R_{2,\rm e}$, and the faster the mass loss,
the  larger the difference $R_2 - R_{2,\rm e}$. Suppose now that the
driving AML rate drops by a large factor on a short time scale. What
will happen? The donor will detach from its Roche lobe because
initially it will continue losing mass and shrink at the same rate as
before while its Roche radius, because of the reduced AML rate will
shrink much more slowly. So mass transfer stops and the star, being
oversized because of previous high mass loss but now without mass loss
contracts towards its thermal equilibrium radius $R_{2,\rm e}$, and
that  on its thermal time scale which is initially shorter than the
time scale on which its Roche radius shrinks. Mass transfer can only
resume when the shrinking Roche radius reaches the stellar radius,
i.e. the latest when $R_{2,{\rm R}} = R_{2, \rm e}$. Once mass transfer  
resumes the binary's orbital period is $P_{\rm l} < P_{\rm u}$. In
other words: the binary has crossed the period range $P_{\rm l}
\leqslant P_{\rm orb} \leqslant P_{\rm u}$ as a detached system. And
because of lacking accretion luminosity, such systems are
intrinsically very faint, fainter even than pre-CVs, and, therefore,
virtually unobservable. A gap in the period distribution can thus
arise if a) a sudden drop of the AML rate causes CVs to detach, b) if
that happens to most of the CVs evolving from $P_{\rm orb} > P_{\rm u}
\rightarrow P_{\rm orb} < P_{\rm u}$, and if c) the values of $P_{\rm u}$ 
and $P_{\rm l}$ are practically the same for all systems going through
a detached phase. 

So far I have not yet addressed the question why the AML rate should
drop by a large factor and that also on a sufficiently short time
scale. The idea behind this proposition is that effective
amplification of magnetic flux via a dynamo and thus efficient AML loss
via  ``magnetic braking'' is strongly tied to the presence of a 
convective envelope and a radiative core in the donor star
(e.g.\ \citet{Spruit&Ritter1983}). Accordingly, it is proposed that
AML via ``magnetic braking'' decreases rapidly when, as a consequence
of ongoing mass loss, the donor eventually becomes fully convective.
The questions of whether that really happens and whether AML via 
``magnetic braking'' stops completely or only partially when the donor
becomes fully convective have remained somewhat controversial to this
day. Qualitative theoretical arguments in favour of the above
proposition have, however, been presented by \citet{T&S1989}. 

In order for the {\em disrupted magnetic braking} proposition to work 
quantitatively the following requirements must be met: AML above the
gap must drive mass transfer at a level of $-\dot M_2 \sim 10^{-9}
M_\odot {\rm yr}^{-1}$. As a result, the donor becomes fully convective
when $M_{\rm conv} \sim 0.2 M_\odot$  and $P_{\rm u} \sim 3^{\rm
  h}$. At that moment, as a consequence  of previous high mass loss,
the stellar radius is larger by about 30\% than in thermal
equilibrium. With the disappearance of the AML from ``magnetic
braking'' the AML loss rate drops by a factor of $\sim 10 - 20$ to
essentially the value due to gravitational radiation alone. After the
detached phase which lasts $\sim 10^9\,{\rm yr}$ mass transfer resumes
with $M_2 = M_{\rm conv} \sim 0.2 M_\odot$, $R_2 = R_{2,\rm e} \sim
0.2 R_\odot$ and $P_{\rm orb} = P_{\rm l} \sim 2^{\rm h}$ at a level of   
$-\dot M_2 \sim 5\,10^{-11} M_\odot {\rm   yr}^{-1}$. Explaining the
gap as a collective phenomenon of CV evolution requires furthermore
that the majority of the donor stars are all of the same type, i.e. MS
stars, and that AML via ``magnetic braking'' yields similar mass
transfer rates in different systems at the same orbital period. This
guarantees that $P_{\rm u}$ and $P_{\rm l}$ are more or less the same
for  all systems and thus the coherence of the phenomenon. 

The fact that the period range of the gap is not empty already
indicates that not all CVs follow the above-described evolution
strictly. There are several reasons for why there may be CVs in the
gap. The most important ones are: 1) a donor mass such that at the end 
of the detached pre-CV evolution the orbital period is $2^{\rm h}
\lesssim P_{\rm orb} \lesssim 3^{\rm h}$ (e.g.\ \citet{Kolb1993},
\citet{DKWG2008}); 2) a donor star which initially was close to the
terminal age MS (see e.g.\ \citet{Ritter1994}), or which is the
artificially evolved remnant of earlier thermal time scale mass
transfer \citep{S&K2002}; 3) reduced ``magnetic braking'' because of
the presence of a strongly magnetized WD (for details see
\citet{LWW1994}).  

At the end of CV evolution the donor star is a very faint brown dwarf.
The WD, in turn, with an effective temperature of typically
$<\,10^4{\rm K}$ is also very faint. And because the mass transfer
rate resulting from gravitational radiation is very small as well,
i.e. $-\dot M_2 \lesssim 10^{-11} M_\odot{\rm   yr}^{-1}$, so is the
resulting accretion luminosity. Thus, such CVs are extremely faint and
inconspicuous objects, and correspondingly difficult to detect. And
though intrinsically about 90\% of all CVs are in this late phase
\citep{Kolb1993} so far only one convincing candidate beyond and far
from the period minimum is known \citep{LDMGSW2006}. The CV graveyard,
as this evolutionary branch is sometimes referred to, is thus largely
hidden from our view.

\begin{acknowledgements}
I would like to thank the organizers of the School of Astrophysics
``Francesco Lucchin'' for their invitation to lecture at this
memorable school. 
\end{acknowledgements}

\bibliographystyle{aa}

\end{document}